\documentclass[aps,pre,amsmath,lengthcheck,superscriptaddress]{revtex4-2}

\usepackage{graphicx}\graphicspath{ {figures/} }
\usepackage{hyperref}
\hypersetup{colorlinks,allcolors=blue,breaklinks}

\newcommand{\be}{\begin{equation}}
\newcommand{\ee}{\end{equation}}
\newcommand{\ba}{\begin{align}}
\newcommand{\ea}{\end{align}}
\newcommand{\bi}{\begin{itemize}}
\newcommand{\ei}{\end{itemize}}

\newcommand{\tr}[1]{\mathrm{tr}\left\{#1\right\}}

\newcommand{\la}{\left\langle}
\newcommand{\ra}{\right\rangle}
\newcommand{\pd}{\partial}

\newcommand{\bla}{bla\\bla\\bla\\bla\\bla}

\newcommand{\mc}[1]{\mathcal{#1}}

\begin{document}

\title{Quantum Ising chain with time-averaged work in linear response theory}

\author{Pierre Naz\'e}
\email{pierre.naze@unesp.br}

\affiliation{\it Universidade Estadual Paulista, 14800-090, Araraquara, S\~ao Paulo, Brazil}

\date{\today}

\begin{abstract}

For systems performing a weakly isothermal process, the decorrelation time dictates how fast the relaxation function decorrelates. However, like many other thermally isolated systems, the transverse-field quantum Ising chain presents an ill-defined decorrelation time. On the other hand, the Kibble-Zurek mechanism uses a heuristic relaxation time to achieve its famous scaling. The problem however of having a well-defined decorrelation time, derived from first principles, agreeing with the Kibble-Zurek mechanism is still open. Such a solution is proposed here by measuring the work using the time-averaged relaxation function of the system, which offers a new and well-defined decorrelation time for thermally isolated systems. I recover with this the Kibble-Zurek mechanism in the finite-time and weak driving regime, and new features in the slowly-varying one. The gain in control over the system in such distinction is desirable for potential applications.

\end{abstract}

\maketitle

\section{Introduction}
\label{sec:intro}

Far-from-equilibrium Thermodynamics has naturally extended its equilibrium counterpart to processes performed at a finite time. Notions of how fast or slow is a process are now pertinent in the thermodynamic analysis, and simple parameters that express such ideas become fundamental. For instance, in the context of weakly isothermal processes, such ``velocity'' parameter is represented using the ratio between two characteristic times: the natural decorrelation timescale of the relaxation function of the system and the inverse of the rate of the process \cite{naze2020compatibility}. In this manner, fast processes occur faster than the decorrelation into equilibrium, while in the slower ones the opposite happens.

However, such a decorrelation timescale is not always well-defined in some systems, which makes thermodynamic analysis very difficult. This is what happens for example to thermally isolated systems performing adiabatic driven processes, that is, a driven process performed without contact with a heat bath. For a variety of systems, such as the quantum harmonic oscillator or Landau-Zener model, one can interpret such timescale as a random one~\cite{naze2023adiabatic}.

The paradigmatic example of the quantum Ising chain, very studied today by its applicability in adiabatic quantum computing or quantum annealing~\cite{deffner2019quantum,Morita2008,Hauke2020,Chakrabarti2023,Hegde2022,Khezri2022,King2022,Soriani2022,Yulianti2022,soriani2022assessing}, has such characteristics. Even though, its phenomenology has been largely elucidated over the years with the formulation of the Kibble-Zurek mechanism~\cite{zurek2005dynamics,del2014universality,yukalov2015realization,zamora2020kibble,damski2005simplest,saito2007kibble,nowak2021quantum,schmitt2022quantum}. In this description, it is used a heuristic relaxation time that will dictate its non-equilibrium effects due to the quantum phase transition that the system passes through when it crosses the critical point. The following question is then established: how can one bring into line such different aspects?

In a recent work~\cite{naze2022kibble}, my co-workers and I have shown that, in the context of finite-time and weak driving processes, the upper bound of the oscillatory decorrelation timescale has the same diverging behavior at the critical point exhibited by the Kibble-Zurek mechanism. In another work~\cite{naze2023adiabatic}, again in the same context, I proposed a solution to capture a decorrelation timescale of thermally isolated systems, where such quantity naturally appears if the time average of the relaxation function is taken. In this case, apparently the decorrelation time is connected to the natural timescale of the thermally isolated system~\cite{naze2023adiabatic}. In this work, I combine both ideas of the two mentioned papers, taking the time average of the relaxation function of the quantum Ising chain and expecting the same diverging behavior of the decorrelation time close to the critical point.

The main results are the following: the time-averaging procedure delivers what was expected and, with it, I establish clear regimes where the process is fast and slow, called finite-time and weak driving regime and slowly-varying one. In this manner, I verify that, in the regime where the process is fast, the system behaves as predicted by the Kibble-Zurek mechanism, having the same impulse window and the same scaling with the rate of the process for the excess work calculated in the impulse part. Also, for the regime where the process is slow, the system presents new features, with a fixed impulse window and a new scaling, now calculated at the excess work in the adiabatic part.

As far as I know, the question of finding precisely a decorrelation time for the quantum Ising model crossing the critical point is unexplored. This is due to the simplification that the Kibble-Zurek mechanism is observed only in the thermodynamic limit, where the relaxation time diverges at the critical point due to the closing gap. Therefore, every process occurring around such a point, with any finite rate, is fast. Establishing thus a difference between fast and slow regimes is in this manner pointless. However, these ``sudden quantum quenches'', in practice, are attenuated due to the system having a finite number of spins~\cite{King2022}. Knowing the decorrelation time for such a finite number regime is therefore of extreme importance to potential applications since determining with clarity what are fast or slow regimes is desirable to have control.

\section{Linear response theory:\\ Excess work}
\label{sec:lrt}

Consider a quantum system with a Hamiltonian $\mc{H}(\lambda(t))$, where $\lambda(t)$ is a time-dependent external parameter. Initially, this system is in contact with a heat bath of temperature $\beta\equiv {(k_B T)}^{-1}$, where $k_B$ is Boltzmann's constant. The system is then decoupled from the heat bath and, during a switching time $\tau$, the external parameter is changed from $\lambda_0$ to $\lambda_0+\delta\lambda$. The average work performed on the system during this process is
\be
W(\tau) \equiv \int_0^\tau \la\pd_{\lambda}\mc{H}(t)\ra\dot{\lambda}(t)dt,
\label{eq:work}
\ee
where $\partial_\lambda$ is the partial derivative for $\lambda$ and the superscripted dot is the total time derivative. The generalized force $\la\pd_{\lambda}\mc{H}(t)\ra$ is calculated using the trace over the density matrix $\rho(t)$
\be
\la A(t)\ra =\tr{A\rho(t)}
\ee
where $A$ is some observable. The density matrix $\rho(t)$ evolves according to Liouville equation
\be
\dot{\rho} =\mc{L}\rho:= -\frac{1}{i\hbar}[\rho,\mc{H}],
\ee
where $\mc{L}$ is the Liouville operator, $[\cdot,\cdot]$ is the commutator and $\rho(0)=\rho_c$ is the initial canonical density matrix. Consider also that the external parameter can be expressed as
\be
\lambda(t) = \lambda_0+g(t)\delta\lambda,
\ee
where to satisfy the initial conditions of the external parameter, the protocol $g(t)$ must satisfy the following boundary conditions
\be
g(0)=0,\quad g(\tau)=1. 
\label{eq:bc}
\ee

Linear response theory aims to express the average of some observable until the first order of some perturbation considering how this perturbation affects the observable and the non-equilibrium density matrix \cite{kubo2012}. In our case, we consider that the parameter considerably does not change during the process, $|g(t)\delta\lambda/\lambda_0|\ll 1$, for all $t\in[0,\tau]$. Using the framework of linear-response theory, the generalized force can be approximated until the first order as
\begin{equation}
\begin{split}
\la\pd_{\lambda}\mc{H}(t)\ra =&\, \la\pd_{\lambda}\mc{H}\ra_0+\delta\lambda\la\pd_{\lambda\lambda}^2\mc{H}\ra_0 g(t)\\
&-\delta\lambda\int_0^t \phi_0(t-t')g(t')dt',
\label{eq:genforce-resp}
\end{split}
\end{equation}
where the $\la\cdot\ra_0$ is the average over the initial canonical density matrix. The quantity $\phi_0(t)$ is the so-called response function, which can be conveniently expressed as the derivative of the relaxation function $\Psi_0(t)$
\be
\phi_0(t) = -\frac{d \Psi_0}{dt},
\label{eq:resprelax}
\ee 
where
\be
\Psi_0(t)=\beta\langle\partial_\lambda\mathcal{H}(t)\partial_\lambda\mathcal{H}(0)\rangle_0+\mathcal{C}
\ee
being the constant $\mathcal{C}$ calculated via the final value theorem \cite{kubo2012}. We define the decorrelation time as the value
\be
\tau_c = \int_0^\infty \frac{\Psi_0(t)}{\Psi_0(0)}dt,
\ee
which measures the time necessary for the relaxation function to decorrelates. In this manner, the generalized force, written in terms of the relaxation function, is
\begin{equation}
\begin{split}
\la\pd_{\lambda}\mc{H}(t)\ra =&\, \la\pd_{\lambda}\mc{H}\ra_0-\delta\lambda\widetilde{\Psi}_0 g(t)\\
&+\delta\lambda\int_0^t \Psi_0(t-t')\dot{g}(t')dt',
\label{eq:genforce-relax}
\end{split}
\end{equation}
where $\widetilde{\Psi}_0(t)\equiv \Psi_0(0)-\la\pd_{\lambda\lambda}^2\mc{H}\ra_0$. Combining Eqs. (\ref{eq:work}) and (\ref{eq:genforce-relax}), the average work performed at the linear response of the generalized force is
\begin{equation}
\begin{split}
W(\tau) = &\, \delta\lambda\la\pd_{\lambda}\mc{H}\ra_0-\frac{\delta\lambda^2}{2}\widetilde{\Psi}_0\\
&+\delta\lambda^2 \int_0^\tau\int_0^t \Psi_0(t-t')\dot{g}(t')\dot{g}(t)dt'dt.
\label{eq:work2}
\end{split}
\end{equation}

We remark that in thermally isolated systems, the work is separated into two contributions: the quasistatic work $W_{\rm qs}$ and the excess work $W_{\rm ex}$. We observe that only the double integral on Eq.~(\ref{eq:work2}) has ``memory'' of the trajectory of $\lambda(t)$. Therefore the other terms are part of the contribution of the quasistatic work. Thus, we can split them as
\be
W_{\rm qs} = \delta\lambda\la\pd_{\lambda}\mc{H}\ra_0-\frac{\delta\lambda^2}{2}\widetilde{\Psi}_0,
\ee  
\begin{equation}
\begin{split}
W_{\text{ex}}(\tau) = \delta\lambda^2 \int_0^\tau\int_0^t \Psi_0(t-t')\dot{g}(t')\dot{g}(t)dt'dt.
\label{eq:wirrder0}
\end{split}
\end{equation}
In particular, the excess work can be rewritten using the symmetry property of the relaxation function, $\Psi(t)=\Psi(-t)$ (see Ref.~\cite{kubo2012}),
\begin{equation}
\begin{split}
W_{\text{ex}}(\tau) = \frac{\delta\lambda^2}{2} \int_0^\tau\int_0^\tau \Psi_0(t-t')\dot{g}(t')\dot{g}(t)dt'dt,
\label{eq:wirrder}
\end{split}
\end{equation}
which holds for systems presenting time-reversal symmetry. We remark that such treatment can be applied to classic systems, by changing the operators to functions, and the commutator by the Poisson bracket \cite{kubo2012}.

\section{Time-averaged excess work}
\label{sec:taew}

Thermally isolated systems performing an adiabatic driven process can be interpreted as having a random decorrelation time \cite{naze2023adiabatic}. Therefore, at each instant of time that the process is performed, the relaxation function changes with it. This is very similar to what happens with systems performing an isothermal process, where the stochastic aspect of the dynamics changes the relaxation function. In this case, we take a stochastic average on the work to correct such an effect. In the case of thermally isolated systems, I have proposed as a solution to this problem in Ref.~\cite{naze2023adiabatic} the following time-averaging
\be
\overline{W}(\tau)=\frac{1}{\tau}\int_0^\tau W(t)dt.
\ee
Such quantity can be measured in the laboratory considering an average in the data set of processes executed in the following way: first, we choose a switching time $\tau$. After, we randomly choose an initial condition from the canonical ensemble and a time $t$ from a uniform distribution, where $0<t<\tau$. Removing the heat bath, we perform the work by changing the external parameter and collecting then its value at the end. The data set produced will furnish, on average, the time-averaged work. 

In the following, I present how time-averaged work can be calculated using linear-response theory and how one can calculate the decorrelation time associated to the system. To do so, we define the idea of time-averaged excess work
\be
\overline{W}_{\rm ex}(\tau) = \frac{1}{\tau}\int_0^\tau W_{\rm ex}(t)dt,
\ee
where $W(\tau)=W_{\rm ex}(\tau)+W_{\rm qs}$.

Now we observe how the time-averaged excess work can be calculated using linear-response theory. In Ref.~\cite{naze2023adiabatic}, I have shown that 
\be
\overline{W}_{\text{ex}}(\tau) = \delta\lambda^2\int_0^\tau\int_0^t \overline{\Psi}_0(t-t')\dot{g}(t)\dot{g}(t')dtdt',
\label{eq:TAexcesswork}
\ee
where
\be
\overline{\Psi}_0(t)=\frac{1}{t}\int_0^t \Psi_0(u)du,
\label{eq:relaxfuncaverage}
\ee
is the time-averaged relaxation function. This means that calculating the time-averaged excess work is the same as calculating the averaged excess work but with a time-averaged relaxation function. Again, this is quite similar to what happens to systems performing isothermal processes, where a stochastic average is taken on the relaxation function. 

Now, when measured with time-averaged work, the thermally isolated system presents an associated decorrelation time. Indeed, the conditions such that linear-response theory is compatible with the Second Law of Thermodynamics are~\cite{naze2020compatibility}
\be
\widetilde{\overline{\Psi}}_0(0)<\infty,\quad \Hat{\Hat{\overline{\Psi}}}_0(\omega)\ge 0,
\ee
where $\widetilde{\cdot}$ and $\Hat{\Hat{\cdot}}$ are respectively the Laplace and Fourier transforms. Therefore, because of this, and analogously to what happens in an isothermal process, we define a new decorrelation time
\be
\overline{\tau}_c := \int_0^\infty\frac{\overline{\Psi}_0(t)}{\overline{\Psi}_0(0)}dt=\frac{\widetilde{\overline{\Psi}}_0(0)}{\overline{\Psi}_0(0)}<\infty.
\label{eq:TArelaxtime}
\ee

It is important to emphasize that the time-average work is a procedure to measure an energy spent during non-equilibrium drivings. It does not say that the system is somehow relaxing, in a sense of dissipation. Indeed, its thermally isolated aspects remain during the procedure of measure. What decorrelates is only the time-average relaxation function. Apparently, the decorrelation time associated is directly connected to the natural timescale of the thermally isolated system~\cite{naze2023adiabatic}.

\section{Kibble-Zurek mechanism}

In what follows, I describe the Kibble-Zurek mechanism and its main characteristics.

\subsection{Phenomenology}

Consider the transverse-field quantum Ising chain, whose Hamiltonian is
\be
\mathcal{H} = -J\sum_{i=1}^{N} \sigma_i^x\sigma_{i+1}^x-\Gamma\sum_{i=1}^{N} \sigma_i^z.  
\label{eq:qim}
\ee
where each one of the $N$ spins has a vector $\vec{\sigma}_i := \sigma_i^x {\bf x}+ \sigma_i^y {\bf y}+ \sigma_i^z {\bf z} $ composed by the Pauli matrices. The parameter $J$ is the coupling energy and $\Gamma$ is the transverse magnetic field.  We assume for simplicity that $N$ is an even number, and the spins obey a periodic boundary condition, $\hat{\vec{\sigma}}_{N+1} = \hat{\vec{\sigma}}_1$. Also, that the system starts with $T=0$.

The Kibble-Zurek mechanism is a phenomenological theory that describes the non-equilibrium dynamics of the transverse-field quantum Ising chain around the critical point $\Gamma=J$. It predicts as well the scaling behavior of observables in the driving rate $1/\tau$ when the system crosses the critical point \cite{zurek2005dynamics,del2014universality,deffner2019quantum}. 

\begin{figure}
\centering
\includegraphics[scale=0.50]{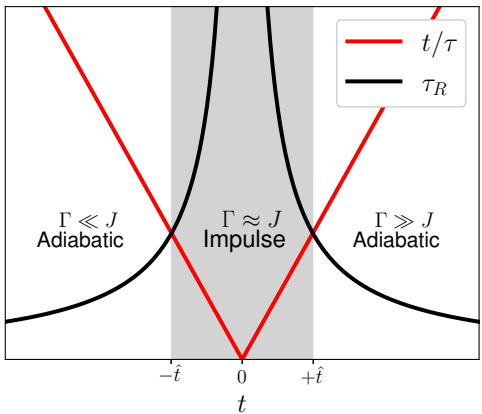}
\caption{Illustration of the Kibble-Zurek mechanism. Far from the critical point, the dynamics of the system are essentially adiabatic, meaning that the system recovers from the defects of the driving faster than the inverse of the driving rate. After the instant $-\hat{t}$ (see Eq.~\eqref{eq:that}), and close to the critical point, the situation changes dramatically. The healing capacity is lost and there is a creation of finite-size magnetic domains.}
\label{fig:1}
\end{figure}

To have a better understanding of this heuristic theory, suppose that the magnetic field $\Gamma$ is driven by a linear protocol
\begin{equation}
\Gamma(t) = J\left|1- r(t)\right|,\quad r(t)=\frac{t}{\tau}.
\end{equation}
Figure \ref{fig:1} illustrates the Kibble-Zurek mechanism. When the system is far enough from the critical point, the dynamics are adiabatic, and the excitations and topological defects heal faster than when they are created. That particular region is called adiabatic. However, when the system approaches the critical point, passing in a particular instant of time $-\hat{t}$, the relaxation time of the system increases dramatically, and the capacity for healing is lost. This phenomenon is manifested by the appearance of finite-sized magnetic domains on the system. That particular region around the critical point is called the impulse one. After the system crosses the impulse region, passing again in a particular instant of time $\hat{t}$, it enters into a new adiabatic region.

The instants when the system passes from adiabatic to impulse regime, which were denoted by $\pm\hat{t}$, are defined as the times when the rate time $r(t)/\dot{r}(t)$ is equal to the rate of the relaxation time $\tau_{R}(t)$. The latter quantity is defined as the time interval for a quantum system to decrease its energy by one energy gap $\Delta(t)$ when it is driven by some protocol at some particular instant $t$
\begin{equation}
\tau_R(t) := \frac{\hbar}{\Delta(t)},
\label{eq:RelaxTimeQM}
\end{equation}
where $\hbar$ is Planck constant. Intuitively, it measures the time for the system to achieve its ground state if the driving is stopped at the instant $t$. In the particular case of the Hamiltonian (\ref{eq:qim}), in the thermodynamic limit, $N\rightarrow\infty$, the gap is
\begin{equation}    
\Delta(t) := 2|J-\Gamma(t)|.
\label{eq:gap}
\end{equation}
This shows that close to the critical point, the gap closes and the relaxation time diverges. This indicates that any driving is fast under such a regime and adiabaticity is lost. To identify the instants where the system starts to present such behavior, one solves the equation 
\be
r(\hat{t})/\dot{r}(\hat{t})=\tau_R(\hat{t}),
\label{eq:thatequation}
\ee
where such instants $\pm\hat{t}$ are
\begin{equation}
\hat{t} = \pm\sqrt{\frac{\hbar \tau}{2 J}},
\label{eq:that}
\end{equation}
which depends on the driving rate by which the system crosses the critical point.

\subsection{Kibble-Zurek scaling}

After describing the phenomenology of the system when it crosses the critical point, the Kibble-Zurek mechanism predicts how observables scale concerning the driving rate when it crosses the critical point. It says that there is a universal exponent $\gamma_{\rm KZ}$ that rules that phenomenon, which depends on the equilibrium critical exponents of the system. In particular, for systems of infinite size, the work $W_{\rm im}$ calculated in the impulse part scales as
\begin{equation}
W_{\rm im}(\tau) \propto \tau^{-\gamma_{\rm KZ}},\quad \gamma_{\rm KZ} = \frac{z\nu}{z\nu +1},
\end{equation}
where $z$ the dynamical critical exponent and $\nu$ the spatial critical exponent. It was assumed that the contributions of excess work due to the adiabatic regions are negligible. In particular, for the transverse-field quantum Ising chain driven in the magnetic field, $z=1$ and $\nu=1$. Therefore
\begin{equation}
\gamma_{\rm KZ} = 1/2.
\end{equation}

\subsection{Weak drivings}

In Ref.~\cite{naze2022kibble}, my co-workers and I have shown that the relaxation function per number of spins for the transverse-field quantum Ising chain is
\be
\Psi_N(t)=\frac{16}{N}\sum_{n=1}^{N/2}\frac{J^2}{\epsilon^3(n)}\sin^2{\left(\left(\frac{2n-1}{N}\right)\pi\right)}\cos{\left(\frac{2\epsilon(n)}{\hbar}t\right)},
\ee
where
\be
\epsilon(n)=2\sqrt{J^2+\Gamma_0^2-2 J \Gamma_0 \cos{\left(\left(\frac{2n-1}{N}\right)\pi\right)}},
\ee
being $\Gamma_0$ the initial value of the magnetic field. Observe that, in this case, the decorrelation time is
\be
\tau_c=\frac{8\hbar}{N}\sum_{n=1}^{N/2}\frac{J^2}{\epsilon^4(n)}\sin^2{\left(\left(\frac{2n-1}{N}\right)\pi\right)}\sin{(\infty)},
\ee
which is ill-defined. To solve this problem, consider the time-averaged relaxation function per number of spins. It will be
\be
\overline{\Psi}_N(t)=\frac{16}{N}\sum_{n=1}^{N/2}\frac{J^2}{\epsilon^3(n)}\sin^2{\left(\left(\frac{2n-1}{N}\right)\pi\right)}{\rm sinc}{\left(\frac{2\epsilon(n)}{\hbar}t\right)},
\label{eq:TApsi}
\ee
where
\be
{\rm sinc}(x) = \frac{\sin{(x)}}{x}.
\ee

Now we can calculate the new decorrelation time and analyze if the system will be affected by the Kibble-Zurek mechanism.

\section{Time-averaged decorrelation time}

\begin{figure}
\centering
\includegraphics[scale=0.50]{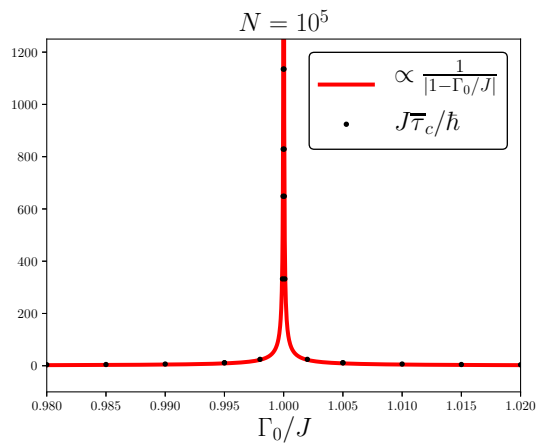}
\caption{Time-averaged decorrelation time according to Eq.~\eqref{eq:TArelaxtime}. It presents a good agreement with the Kibble-Zurek mechanism prediction. It was used $N=10^5$.}
\label{fig:2}
\end{figure}

Given the time-averaged relaxation function per number of spins~\eqref{eq:TApsi}, and using Eq.~\eqref{eq:TArelaxtime}, the time-averaged decorrelation time will be
\be
\overline{\tau}_c(\Gamma_0) = \frac{\sum_{i=1}^{N/2}\frac{\pi\hbar}{\epsilon^4(n)}\sin^2{\left(\left(\frac{2n-1}{N}\right)\pi\right)}}{\sum_{i=1}^{N/2}\frac{4}{\epsilon^3(n)}\sin^2{\left(\left(\frac{2n-1}{N}\right)\pi\right)}},
\ee
which is naturally measured in units of $\hbar/J$. For a large number of spins, $N\gg 1$, one has
\be
\lim_{N\gg 1} \overline{\tau}_c(\Gamma_0)\approx \frac{2\pi\hbar|J-\Gamma_0|^{-4}\sin^2{\left(\pi\right)}[(1+N/2)N/4]}{16|J-\Gamma_0|^{-3}\sin^2{\left(\pi\right)}[(1+N/2)N/4]},
\ee
which leads to
\be
\lim_{N\gg 1} \overline{\tau}_c(\Gamma_0) \propto \frac{\hbar}{|J-\Gamma_0|}.
\label{eq:decorrelationtime}
\ee
Observe that such decorrelation time, in the thermodynamic limit, has the same explanation of the diverging behavior of the heuristic relaxation time of the Kibble-Zurek mechanism: the closing gap near the critical point. Figure~\ref{fig:2} shows this same diverging behavior as well. Therefore, we have now a decorrelation time, derived from first principles, agreeing with the Kibble-Zurek mechanism heuristic explanation.

\begin{figure}
\centering
\includegraphics[scale=0.50]{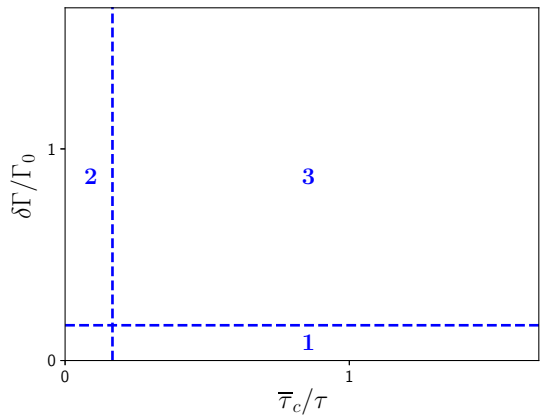}
\caption{Diagram of non-equilibrium regions. Region 1 corresponds to finite-time and weak processes, region 2 to slowly-varying processes, and region 3 to far-from-equilibrium processes. Linear-response theory can describe regions 1 and 2.}
\label{fig:3}
\end{figure}

Now one can identify the non-equilibrium regimes of the process. This is done using the ratio between the decorrelation time and switching time, which informs how fast the driving is performed, and the ratio $\delta\Gamma/\Gamma_0$, which informs how strong the process is. One can create a diagram of non-equilibrium regions illustrating that. See Fig.~\ref{fig:3}. In region 1, the so-called finite-time and weak processes, the ratio $\delta\Gamma/\Gamma_0\ll 1$, while $\overline{\tau}_c/\tau$ is arbitrary. By contrast, in region 2, the so-called slowly-varying processes, the ratio $\delta\Gamma/\Gamma_0$ is arbitrary, while $\overline{\tau}_c/\tau\ll 1$. In region 3, the so-called arbitrarily far-from-equilibrium processes, both ratios are arbitrary. Linear-response theory can calculate the time-averaged excess works of regions 1 and 2 \cite{naze2022optimal}.

Indeed, for region 1, the time-averaged excess work per number of spins is given by
\be
\overline{W}_{\text{ex}}^{(1)}(\tau) = \int_0^\tau\int_0^t \overline{\Psi}_N(t-t')\dot{\Gamma}(t)\dot{\Gamma}(t')dtdt',
\label{eq:TAexcesswork1}
\ee
For region 2, using the asymptotic approximation of decorrelation of the time-averaged relaxation function per number of spins
\cite{naze2022optimal}
\be
\lim_{\overline{\tau}_c/\tau\ll 1}\overline{\Psi}_N(t) = 2\overline{\tau_c} \overline{\Psi}_N(0)\delta(t),
\label{eq:approximationSL}
\ee
the time-averaged excess work per number of spins will be
\be
\overline{W}_{\rm ex}^{(2)}(\tau) = \int_0^\tau \overline{\tau}_c[\Gamma(t)]\overline{\chi}_N[\Gamma(t)]\Gamma(t)^2dt,
\label{eq:TAexcesswork2}
\ee
where 
\be
\overline{\chi}_N[\Gamma_0] = \overline{\Psi}_N(0).
\ee
An important observation is necessary to be made: the distinction between systems of finite size and infinite size. In the first case, the regime of region 2 is well-defined, while in the second one, only the one of region 1 exists. Indeed, Eq.~\eqref{eq:TAexcesswork2} is only valid for $\tau\gg\tau_c(\Gamma(t))$, whose highest value occurs at $\Gamma=J$. For systems of finite size, such value is finite, while for infinite size, is not. Therefore, we can only find suitable switching times for the regime of region 2 in the first case. Also, since the Kibble-Zurek mechanism describes systems of infinite size, every process is been performed outside the slowly-varying processes regime, that is, is a fast process. Such a process is what is usually called ``sudden quantum quench''. 

As Ref.~\cite{naze2022kibble} has presented, the range of validity of numerical precision of linear-response theory of the regime of region 1 is only valid for systems of finite size and very small perturbations, although it predicts the Kibble-Zurek mechanism qualitative behaviors. We assume that such behavior will happen here in the time-averaged context. Finally, in our simulations, we are going to work only with systems of finite size, which will allow us to detect the effects of the Kibble-Zurek mechanism in finite-time and weak processes and observe the new features in the regime of slowly-varying processes. 

\section{Adiabatic and impulse regions}

\begin{figure}
\centering
\includegraphics[scale=0.50]{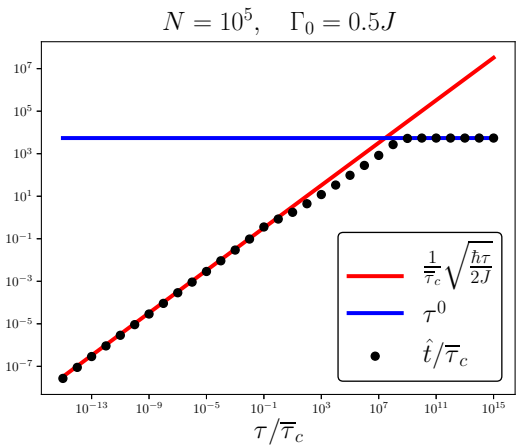}
\caption{Illustration of $\hat{t}/\overline{\tau}_c$ for different $\tau/\overline{\tau}_c$. For $\tau\ll\overline{\tau}_c$, $\hat{t}\propto\tau^{1/2}$, as predicted by Kibble-Zurek mechanism. For $\tau\gg\overline{\tau}_c$, $\hat{t}$ achieves a plateau. It was used $N=10^5, \Gamma_0=0.5J$.}
\label{fig:4}
\end{figure}

To evaluate if the time-averaged excess work per number of spins calculated in the impulse region is much bigger than its adiabatic counterparts, we have to calculate first the interval of time where this impulsive part occurs. To do so, we evaluate Eq.~\eqref{eq:thatequation} with our analogous quantities. In this case, considering the following linear driving 
\be
\Gamma(t)=(J-\Gamma_0)+2\Gamma_0 t/\tau,
\label{eq:lineardriving}
\ee
we need to solve
\be
\frac{\Gamma(t)-J}{\dot{\Gamma}(t)}=\left|\frac{\tau}{2}-t\right| = \overline{\tau}_c(\Gamma(t)).
\label{eq:TAthatequation}
\ee
The graphic of $\hat{t}/\overline{\tau}_c$, plotted against $\tau/\overline{\tau}_c$, is depicted in Fig.~\ref{fig:4}. It was used $N=10^5$ and $\Gamma_0=0.5 J$. As predicted by Kibble-Zurek mechanism, for $\tau\le\tau_c(\Gamma_0)$, $\hat{t} = \sqrt{\hbar\tau/2J}$. Also, for $\tau\gg\tau_c$, $\hat{t}$ achieves a plateau. This means that the window for the impulse part remains the same size, even though the duration of the process becomes larger for slower rates.

The next step is to calculate the impulse and adiabatic time-averaged excess work per number of spins in both regimes. For finite-time and weak processes, and $\tau\ll\overline{\tau}_c(\Gamma_0)$, we have in the impulse part
\be
\overline{W}_{\rm im}^{(1)}(\tau)=\int_{\tau/2-\sqrt{\hbar\tau/2J}}^{\tau/2+\sqrt{\hbar\tau/2J}}\int_0^t \overline{\Psi}_N(t-t')\dot{\Gamma}(t)\dot{\Gamma}(t')dtdt',
\label{eq:weximpulse2}
\ee
while its adiabatic counterpart is 
\begin{equation}
\overline{W}_{\rm ad}^{(1)}(\tau)=\overline{W}^{(1)}(\tau)-\overline{W}_{\rm im}^{(1)}(\tau).
\end{equation}
For slowly-varying processes, and $\tau\gg\overline{\tau}_c(J)$, we have in the impulse part
\be
\overline{W}_{\rm im}^{(2)}(\tau)=\int_{\tau/2-c}^{\tau/2+c} \overline{\tau}_c(\Gamma(t))\overline{\chi}(\Gamma(t))\dot{\Gamma}(t)^2dt,
\label{eq:weximpulse2}
\ee
while its adiabatic counterpart is 
\be
\overline{W}_{\rm ad}^{(2)}(\tau)=\overline{W}^{(2)}(\tau)-\overline{W}_{\rm im}^{(2)}(\tau).
\ee
Here, the constant $c$ will be evaluated from the solution of Eq. \eqref{eq:TAthatequation}.

From Fig.~\ref{fig:4} it is possible to evaluate the proportion between the adiabatic and impulse parts in regime 1, for $\tau\ll\overline{\tau}_c(\Gamma_0)$, $N=10^5$ and $\Gamma_0=0.5J$. For instance, for $\tau=0.01\overline{\tau}_c$, $\hat{t}=0.1\overline{\tau}_c$, which indicates that the whole driving occurs at the impulse region. Therefore, the mentioned proportion is null. Indeed, as predicted by the Kibble-Zurek mechanism, the adiabatic part can be neglected without so much loss in the final result of the time-averaged excess work. On the other hand, for $\tau\gg\overline{\tau}_c$, the proportion diverges since the adiabatic part goes to the quasistatic work and the impulse part to zero. In this situation, is not more useful to calculate the impulse part, which is null, but the adiabatic part only.  

\section{Kibble-Zurek scalings}

\begin{figure}[!ht]
\centering
\includegraphics[scale=0.50]{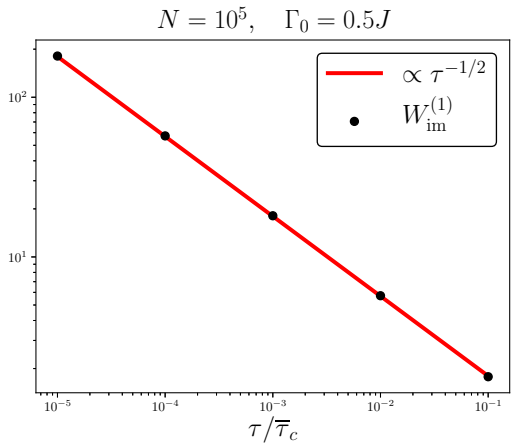}
\caption{Scale $\tau^{-1/2}$ for time-averaged excess work per number of spins in the impulse part for the regime of region 1. It agrees with the prediction of the Kibble-Zurek mechanism. It was used $N=10^5$ and $\Gamma_0=0.5J$.}
\label{fig:5}
\end{figure}

\begin{figure}
\centering
\includegraphics[scale=0.50]{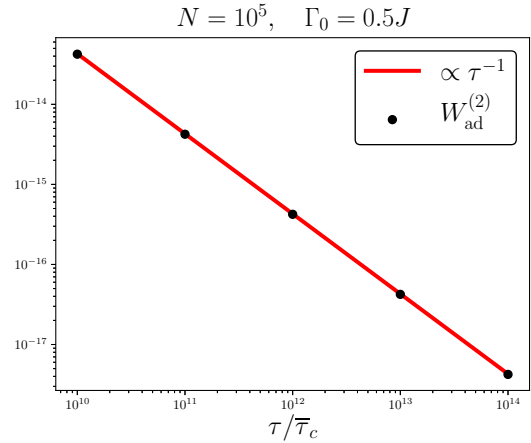}
\caption{Scale $\tau^{-1}$ for time-averaged excess work per number of spins in the adiabatic part for the regime of region 2. It was used $N=10^5$ and $\Gamma_0=0.5J$.}
\label{fig:6}
\end{figure}

Using the linear protocol \eqref{eq:lineardriving}, we explore the rate scalings of $\overline{W}_{\rm im}^{(1)}(\tau)$ and $\overline{W}_{\rm ad}^{(2)}(\tau)$ respectively in the conditions of  $\tau\ll\overline{\tau}_c(\Gamma_0)$ and $\tau\gg\overline{\tau}_c(J)$. In this manner, for the first case, Fig.~\ref{fig:5} depicts the scaling $\tau^{-1/2}$. Again, such an effect is predicted by the Kibble-Zurek mechanism. In the second case, Fig.~\ref{fig:6} depicts the scaling $\tau^{-1}$. Such behavior is in agreement with the scaling predicted in Ref.~\cite{naze2020compatibility}, since the work calculated in the adiabatic part is practically done in the whole driving range. It agrees also with the prediction of Ref.~\cite{deffner2017}, where the same approximation for slowly-varying processes framework was used.

It is interesting to remark that, with our new framework about decorrelation time and out-of-equilibrium regimes, the scale $\tau^{-1}$ measured in Ref.~\cite{naze2022kibble} was made considering the impulse excess work measured in $\tau\gtrsim\tau_c(J)$. Evaluating the scale in the same range of switching times with our new results of $\hat{t}$, it presents a deviation to $\tau^{-1.1}$. This result shows that the assumption $\hat{t}=\sqrt{\hbar\tau/2J}$ made in the previous work is not so good for this case (see Fig.~\ref{fig:4}). Here, I assumed that the normal and time-averaged cases should have the same scaling.

\section{Final remarks}

Using the time-averaged relaxation function, I calculated a well-defined decorrelation time for the transverse-field quantum Ising chain. For a large number of spins, this decorrelation time presents a diverging behavior close to the critical point, which was shown to have the same closing gap explanation that the heuristic relaxation time presents. Also, with such decorrelation time, two different regimes, the finite-time and weak regime, and the slowly-varying one, were plenty established. In the first regime, I found the effects predicted by the Kibble-Zurek mechanism, while, in the second one, two interesting behaviors: a $\hat{t}$ achieving a plateau and a time-averaged excess work calculated in the adiabatic part of the process scaling with $\tau^{-1}$. This is the first step to see that the time-averaged work is indeed a useful quantity to measure the excess work in potential applications since it offers in this case more control over the regimes where the driving system operates. Future research will be done in this direction.

\bibliography{KZTA.bib}

\begin{thebibliography}{24}%
\makeatletter
\providecommand \@ifxundefined [1]{%
 \@ifx{#1\undefined}
}%
\providecommand \@ifnum [1]{%
 \ifnum #1\expandafter \@firstoftwo
 \else \expandafter \@secondoftwo
 \fi
}%
\providecommand \@ifx [1]{%
 \ifx #1\expandafter \@firstoftwo
 \else \expandafter \@secondoftwo
 \fi
}%
\providecommand \natexlab [1]{#1}%
\providecommand \enquote  [1]{``#1''}%
\providecommand \bibnamefont  [1]{#1}%
\providecommand \bibfnamefont [1]{#1}%
\providecommand \citenamefont [1]{#1}%
\providecommand \href@noop [0]{\@secondoftwo}%
\providecommand \href [0]{\begingroup \@sanitize@url \@href}%
\providecommand \@href[1]{\@@startlink{#1}\@@href}%
\providecommand \@@href[1]{\endgroup#1\@@endlink}%
\providecommand \@sanitize@url [0]{\catcode `\\12\catcode `\$12\catcode
  `\&12\catcode `\#12\catcode `\^12\catcode `\_12\catcode `\%12\relax}%
\providecommand \@@startlink[1]{}%
\providecommand \@@endlink[0]{}%
\providecommand \url  [0]{\begingroup\@sanitize@url \@url }%
\providecommand \@url [1]{\endgroup\@href {#1}{\urlprefix }}%
\providecommand \urlprefix  [0]{URL }%
\providecommand \Eprint [0]{\href }%
\providecommand \doibase [0]{https://doi.org/}%
\providecommand \selectlanguage [0]{\@gobble}%
\providecommand \bibinfo  [0]{\@secondoftwo}%
\providecommand \bibfield  [0]{\@secondoftwo}%
\providecommand \translation [1]{[#1]}%
\providecommand \BibitemOpen [0]{}%
\providecommand \bibitemStop [0]{}%
\providecommand \bibitemNoStop [0]{.\EOS\space}%
\providecommand \EOS [0]{\spacefactor3000\relax}%
\providecommand \BibitemShut  [1]{\csname bibitem#1\endcsname}%
\let\auto@bib@innerbib\@empty
\bibitem [{\citenamefont {Naz{\'e}}\ and\ \citenamefont
  {Bonan{\c{c}}a}(2020)}]{naze2020compatibility}%
  \BibitemOpen
  \bibfield  {author} {\bibinfo {author} {\bibfnamefont {P.}~\bibnamefont
  {Naz{\'e}}}\ and\ \bibinfo {author} {\bibfnamefont {M.~V.~S.}\ \bibnamefont
  {Bonan{\c{c}}a}},\ }\href
  {https://iopscience.iop.org/article/10.1088/1742-5468/ab54ba/meta?casa_token=g65yJG1-SjcAAAAA:JGm2sIKbQ0tYJ-bdvw8Jz75bB3ZMiKL8-YMtvQeZg_v0yB7hEQAzAqET6giEuUDB9rdZahI7Zlg}
  {\bibfield  {journal} {\bibinfo  {journal} {Journal of Statistical Mechanics:
  Theory and Experiment}\ }\textbf {\bibinfo {volume} {2020}},\ \bibinfo
  {pages} {013206} (\bibinfo {year} {2020})}\BibitemShut {NoStop}%
\bibitem [{\citenamefont {Naz\'e}(2023)}]{naze2023adiabatic}%
  \BibitemOpen
  \bibfield  {author} {\bibinfo {author} {\bibfnamefont {P.}~\bibnamefont
  {Naz\'e}},\ }\href {https://doi.org/10.1103/PhysRevE.107.064114} {\bibfield
  {journal} {\bibinfo  {journal} {Phys. Rev. E}\ }\textbf {\bibinfo {volume}
  {107}},\ \bibinfo {pages} {064114} (\bibinfo {year} {2023})}\BibitemShut
  {NoStop}%
\bibitem [{\citenamefont {Deffner}\ and\ \citenamefont
  {Campbell}(2019)}]{deffner2019quantum}%
  \BibitemOpen
  \bibfield  {author} {\bibinfo {author} {\bibfnamefont {S.}~\bibnamefont
  {Deffner}}\ and\ \bibinfo {author} {\bibfnamefont {S.}~\bibnamefont
  {Campbell}},\ }\href@noop {} {\emph {\bibinfo {title} {Quantum
  Thermodynamics: An introduction to the thermodynamics of quantum
  information}}}\ (\bibinfo  {publisher} {Morgan \& Claypool Publishers},\
  \bibinfo {year} {2019})\BibitemShut {NoStop}%
\bibitem [{\citenamefont {Morita}\ and\ \citenamefont
  {Nishimori}(2008)}]{Morita2008}%
  \BibitemOpen
  \bibfield  {author} {\bibinfo {author} {\bibfnamefont {S.}~\bibnamefont
  {Morita}}\ and\ \bibinfo {author} {\bibfnamefont {H.}~\bibnamefont
  {Nishimori}},\ }\bibfield  {journal} {\bibinfo  {journal} {Journal of
  Mathematical Physics}\ }\textbf {\bibinfo {volume} {49}},\ \href
  {https://doi.org/10.1063/1.2995837} {10.1063/1.2995837} (\bibinfo {year}
  {2008})\BibitemShut {NoStop}%
\bibitem [{\citenamefont {Hauke}\ \emph {et~al.}(2020)\citenamefont {Hauke},
  \citenamefont {Katzgraber}, \citenamefont {Lechner}, \citenamefont
  {Nishimori},\ and\ \citenamefont {Oliver}}]{Hauke2020}%
  \BibitemOpen
  \bibfield  {author} {\bibinfo {author} {\bibfnamefont {P.}~\bibnamefont
  {Hauke}}, \bibinfo {author} {\bibfnamefont {H.~G.}\ \bibnamefont
  {Katzgraber}}, \bibinfo {author} {\bibfnamefont {W.}~\bibnamefont {Lechner}},
  \bibinfo {author} {\bibfnamefont {H.}~\bibnamefont {Nishimori}},\ and\
  \bibinfo {author} {\bibfnamefont {W.~D.}\ \bibnamefont {Oliver}},\ }\bibfield
   {journal} {\bibinfo  {journal} {Reports on Progress in Physics}\ }\textbf
  {\bibinfo {volume} {83}},\ \href {https://doi.org/10.1088/1361-6633/AB85B8}
  {10.1088/1361-6633/AB85B8} (\bibinfo {year} {2020})\BibitemShut {NoStop}%
\bibitem [{\citenamefont {Chakrabarti}\ \emph {et~al.}(2023)\citenamefont
  {Chakrabarti}, \citenamefont {Leschke}, \citenamefont {Ray}, \citenamefont
  {Shirai},\ and\ \citenamefont {Tanaka}}]{Chakrabarti2023}%
  \BibitemOpen
  \bibfield  {author} {\bibinfo {author} {\bibfnamefont {B.~K.}\ \bibnamefont
  {Chakrabarti}}, \bibinfo {author} {\bibfnamefont {H.}~\bibnamefont
  {Leschke}}, \bibinfo {author} {\bibfnamefont {P.}~\bibnamefont {Ray}},
  \bibinfo {author} {\bibfnamefont {T.}~\bibnamefont {Shirai}},\ and\ \bibinfo
  {author} {\bibfnamefont {S.}~\bibnamefont {Tanaka}},\ }\bibfield  {journal}
  {\bibinfo  {journal} {Philosophical Transactions of the Royal Society A:
  Mathematical, Physical and Engineering Sciences}\ }\textbf {\bibinfo {volume}
  {381}},\ \href {https://doi.org/10.1098/RSTA.2021.0419}
  {10.1098/RSTA.2021.0419} (\bibinfo {year} {2023})\BibitemShut {NoStop}%
\bibitem [{\citenamefont {Hegde}\ \emph {et~al.}(2022)\citenamefont {Hegde},
  \citenamefont {Passarelli}, \citenamefont {Scocco},\ and\ \citenamefont
  {Lucignano}}]{Hegde2022}%
  \BibitemOpen
  \bibfield  {author} {\bibinfo {author} {\bibfnamefont {P.~R.}\ \bibnamefont
  {Hegde}}, \bibinfo {author} {\bibfnamefont {G.}~\bibnamefont {Passarelli}},
  \bibinfo {author} {\bibfnamefont {A.}~\bibnamefont {Scocco}},\ and\ \bibinfo
  {author} {\bibfnamefont {P.}~\bibnamefont {Lucignano}},\ }\href
  {https://doi.org/10.1103/PHYSREVA.105.012612/FIGURES/11/MEDIUM} {\bibfield
  {journal} {\bibinfo  {journal} {Physical Review A}\ }\textbf {\bibinfo
  {volume} {105}},\ \bibinfo {pages} {012612} (\bibinfo {year}
  {2022})}\BibitemShut {NoStop}%
\bibitem [{\citenamefont {Khezri}\ \emph {et~al.}(2022)\citenamefont {Khezri},
  \citenamefont {Dai}, \citenamefont {Yang}, \citenamefont {Albash},
  \citenamefont {Lupascu},\ and\ \citenamefont {Lidar}}]{Khezri2022}%
  \BibitemOpen
  \bibfield  {author} {\bibinfo {author} {\bibfnamefont {M.}~\bibnamefont
  {Khezri}}, \bibinfo {author} {\bibfnamefont {X.}~\bibnamefont {Dai}},
  \bibinfo {author} {\bibfnamefont {R.}~\bibnamefont {Yang}}, \bibinfo {author}
  {\bibfnamefont {T.}~\bibnamefont {Albash}}, \bibinfo {author} {\bibfnamefont
  {A.}~\bibnamefont {Lupascu}},\ and\ \bibinfo {author} {\bibfnamefont {D.~A.}\
  \bibnamefont {Lidar}},\ }\href
  {https://doi.org/10.1103/PHYSREVAPPLIED.17.044005/FIGURES/8/MEDIUM}
  {\bibfield  {journal} {\bibinfo  {journal} {Physical Review Applied}\
  }\textbf {\bibinfo {volume} {17}},\ \bibinfo {pages} {044005} (\bibinfo
  {year} {2022})}\BibitemShut {NoStop}%
\bibitem [{\citenamefont {King}\ \emph {et~al.}(2022)\citenamefont {King},
  \citenamefont {Suzuki}, \citenamefont {Raymond}, \citenamefont {Zucca},
  \citenamefont {Lanting}, \citenamefont {Altomare}, \citenamefont {Berkley},
  \citenamefont {Ejtemaee}, \citenamefont {Hoskinson}, \citenamefont {Huang},
  \citenamefont {Ladizinsky}, \citenamefont {MacDonald}, \citenamefont
  {Marsden}, \citenamefont {Oh}, \citenamefont {Poulin-Lamarre}, \citenamefont
  {Reis}, \citenamefont {Rich}, \citenamefont {Sato}, \citenamefont
  {Whittaker}, \citenamefont {Yao}, \citenamefont {Harris}, \citenamefont
  {Lidar}, \citenamefont {Nishimori},\ and\ \citenamefont {Amin}}]{King2022}%
  \BibitemOpen
  \bibfield  {author} {\bibinfo {author} {\bibfnamefont {A.~D.}\ \bibnamefont
  {King}}, \bibinfo {author} {\bibfnamefont {S.}~\bibnamefont {Suzuki}},
  \bibinfo {author} {\bibfnamefont {J.}~\bibnamefont {Raymond}}, \bibinfo
  {author} {\bibfnamefont {A.}~\bibnamefont {Zucca}}, \bibinfo {author}
  {\bibfnamefont {T.}~\bibnamefont {Lanting}}, \bibinfo {author} {\bibfnamefont
  {F.}~\bibnamefont {Altomare}}, \bibinfo {author} {\bibfnamefont {A.~J.}\
  \bibnamefont {Berkley}}, \bibinfo {author} {\bibfnamefont {S.}~\bibnamefont
  {Ejtemaee}}, \bibinfo {author} {\bibfnamefont {E.}~\bibnamefont {Hoskinson}},
  \bibinfo {author} {\bibfnamefont {S.}~\bibnamefont {Huang}}, \bibinfo
  {author} {\bibfnamefont {E.}~\bibnamefont {Ladizinsky}}, \bibinfo {author}
  {\bibfnamefont {A.~J.}\ \bibnamefont {MacDonald}}, \bibinfo {author}
  {\bibfnamefont {G.}~\bibnamefont {Marsden}}, \bibinfo {author} {\bibfnamefont
  {T.}~\bibnamefont {Oh}}, \bibinfo {author} {\bibfnamefont {G.}~\bibnamefont
  {Poulin-Lamarre}}, \bibinfo {author} {\bibfnamefont {M.}~\bibnamefont
  {Reis}}, \bibinfo {author} {\bibfnamefont {C.}~\bibnamefont {Rich}}, \bibinfo
  {author} {\bibfnamefont {Y.}~\bibnamefont {Sato}}, \bibinfo {author}
  {\bibfnamefont {J.~D.}\ \bibnamefont {Whittaker}}, \bibinfo {author}
  {\bibfnamefont {J.}~\bibnamefont {Yao}}, \bibinfo {author} {\bibfnamefont
  {R.}~\bibnamefont {Harris}}, \bibinfo {author} {\bibfnamefont {D.~A.}\
  \bibnamefont {Lidar}}, \bibinfo {author} {\bibfnamefont {H.}~\bibnamefont
  {Nishimori}},\ and\ \bibinfo {author} {\bibfnamefont {M.~H.}\ \bibnamefont
  {Amin}},\ }\href {https://doi.org/10.1038/s41567-022-01741-6} {\bibfield
  {journal} {\bibinfo  {journal} {Nature Physics 2022 18:11}\ }\textbf
  {\bibinfo {volume} {18}},\ \bibinfo {pages} {1324} (\bibinfo {year}
  {2022})}\BibitemShut {NoStop}%
\bibitem [{\citenamefont {Soriani}\ \emph
  {et~al.}(2022{\natexlab{a}})\citenamefont {Soriani}, \citenamefont
  {Naz{\'{e}}}, \citenamefont {Bonan{\c{c}}a}, \citenamefont {Gardas},\ and\
  \citenamefont {Deffner}}]{Soriani2022}%
  \BibitemOpen
  \bibfield  {author} {\bibinfo {author} {\bibfnamefont {A.}~\bibnamefont
  {Soriani}}, \bibinfo {author} {\bibfnamefont {P.}~\bibnamefont {Naz{\'{e}}}},
  \bibinfo {author} {\bibfnamefont {M.~V.}\ \bibnamefont {Bonan{\c{c}}a}},
  \bibinfo {author} {\bibfnamefont {B.}~\bibnamefont {Gardas}},\ and\ \bibinfo
  {author} {\bibfnamefont {S.}~\bibnamefont {Deffner}},\ }\href
  {https://doi.org/10.1103/PhysRevA.105.042423} {\bibfield  {journal} {\bibinfo
   {journal} {Physical Review A}\ }\textbf {\bibinfo {volume} {105}},\ \bibinfo
  {pages} {42423} (\bibinfo {year} {2022}{\natexlab{a}})}\BibitemShut {NoStop}%
\bibitem [{\citenamefont {Yulianti}\ and\ \citenamefont
  {Surendro}(2022)}]{Yulianti2022}%
  \BibitemOpen
  \bibfield  {author} {\bibinfo {author} {\bibfnamefont {L.~P.}\ \bibnamefont
  {Yulianti}}\ and\ \bibinfo {author} {\bibfnamefont {K.}~\bibnamefont
  {Surendro}},\ }\href {https://doi.org/10.1109/ACCESS.2022.3188117} {\bibfield
   {journal} {\bibinfo  {journal} {IEEE Access}\ }\textbf {\bibinfo {volume}
  {10}},\ \bibinfo {pages} {73156} (\bibinfo {year} {2022})}\BibitemShut
  {NoStop}%
\bibitem [{\citenamefont {Soriani}\ \emph
  {et~al.}(2022{\natexlab{b}})\citenamefont {Soriani}, \citenamefont
  {Naz{\'e}}, \citenamefont {Bonan{\c{c}}a}, \citenamefont {Gardas},\ and\
  \citenamefont {Deffner}}]{soriani2022assessing}%
  \BibitemOpen
  \bibfield  {author} {\bibinfo {author} {\bibfnamefont {A.}~\bibnamefont
  {Soriani}}, \bibinfo {author} {\bibfnamefont {P.}~\bibnamefont {Naz{\'e}}},
  \bibinfo {author} {\bibfnamefont {M.~V.}\ \bibnamefont {Bonan{\c{c}}a}},
  \bibinfo {author} {\bibfnamefont {B.}~\bibnamefont {Gardas}},\ and\ \bibinfo
  {author} {\bibfnamefont {S.}~\bibnamefont {Deffner}},\ }\href
  {https://journals.aps.org/pra/abstract/10.1103/PhysRevA.105.052442}
  {\bibfield  {journal} {\bibinfo  {journal} {Physical Review A}\ }\textbf
  {\bibinfo {volume} {105}},\ \bibinfo {pages} {052442} (\bibinfo {year}
  {2022}{\natexlab{b}})}\BibitemShut {NoStop}%
\bibitem [{\citenamefont {Zurek}\ \emph {et~al.}(2005)\citenamefont {Zurek},
  \citenamefont {Dorner},\ and\ \citenamefont {Zoller}}]{zurek2005dynamics}%
  \BibitemOpen
  \bibfield  {author} {\bibinfo {author} {\bibfnamefont {W.~H.}\ \bibnamefont
  {Zurek}}, \bibinfo {author} {\bibfnamefont {U.}~\bibnamefont {Dorner}},\ and\
  \bibinfo {author} {\bibfnamefont {P.}~\bibnamefont {Zoller}},\ }\href
  {https://journals.aps.org/prl/abstract/10.1103/PhysRevLett.95.105701}
  {\bibfield  {journal} {\bibinfo  {journal} {Physical review letters}\
  }\textbf {\bibinfo {volume} {95}},\ \bibinfo {pages} {105701} (\bibinfo
  {year} {2005})}\BibitemShut {NoStop}%
\bibitem [{\citenamefont {Del~Campo}\ and\ \citenamefont
  {Zurek}(2014)}]{del2014universality}%
  \BibitemOpen
  \bibfield  {author} {\bibinfo {author} {\bibfnamefont {A.}~\bibnamefont
  {Del~Campo}}\ and\ \bibinfo {author} {\bibfnamefont {W.~H.}\ \bibnamefont
  {Zurek}},\ }\href
  {https://www.worldscientific.com/doi/abs/10.1142/S0217751X1430018X?casa_token=dxdIE4XU5esAAAAA:HqgcjD5X56aJpV7uxGBCNem68ujh0Jp74-mGAxCzRaOw5rJKk4OLF0ptadmqKZFBbGyj6ammURd9zw}
  {\bibfield  {journal} {\bibinfo  {journal} {International Journal of Modern
  Physics A}\ }\textbf {\bibinfo {volume} {29}},\ \bibinfo {pages} {1430018}
  (\bibinfo {year} {2014})}\BibitemShut {NoStop}%
\bibitem [{\citenamefont {Yukalov}\ \emph {et~al.}(2015)\citenamefont
  {Yukalov}, \citenamefont {Novikov},\ and\ \citenamefont
  {Bagnato}}]{yukalov2015realization}%
  \BibitemOpen
  \bibfield  {author} {\bibinfo {author} {\bibfnamefont {V.}~\bibnamefont
  {Yukalov}}, \bibinfo {author} {\bibfnamefont {A.}~\bibnamefont {Novikov}},\
  and\ \bibinfo {author} {\bibfnamefont {V.~S.}\ \bibnamefont {Bagnato}},\
  }\href {https://www.sciencedirect.com/science/article/pii/S0375960115002066}
  {\bibfield  {journal} {\bibinfo  {journal} {Physics Letters A}\ }\textbf
  {\bibinfo {volume} {379}},\ \bibinfo {pages} {1366} (\bibinfo {year}
  {2015})}\BibitemShut {NoStop}%
\bibitem [{\citenamefont {Zamora}\ \emph {et~al.}(2020)\citenamefont {Zamora},
  \citenamefont {Dagvadorj}, \citenamefont {Comaron}, \citenamefont
  {Carusotto}, \citenamefont {Proukakis},\ and\ \citenamefont
  {Szyma{\'n}ska}}]{zamora2020kibble}%
  \BibitemOpen
  \bibfield  {author} {\bibinfo {author} {\bibfnamefont {A.}~\bibnamefont
  {Zamora}}, \bibinfo {author} {\bibfnamefont {G.}~\bibnamefont {Dagvadorj}},
  \bibinfo {author} {\bibfnamefont {P.}~\bibnamefont {Comaron}}, \bibinfo
  {author} {\bibfnamefont {I.}~\bibnamefont {Carusotto}}, \bibinfo {author}
  {\bibfnamefont {N.}~\bibnamefont {Proukakis}},\ and\ \bibinfo {author}
  {\bibfnamefont {M.}~\bibnamefont {Szyma{\'n}ska}},\ }\href
  {https://journals.aps.org/prl/abstract/10.1103/PhysRevLett.125.095301}
  {\bibfield  {journal} {\bibinfo  {journal} {Physical Review Letters}\
  }\textbf {\bibinfo {volume} {125}},\ \bibinfo {pages} {095301} (\bibinfo
  {year} {2020})}\BibitemShut {NoStop}%
\bibitem [{\citenamefont {Damski}(2005)}]{damski2005simplest}%
  \BibitemOpen
  \bibfield  {author} {\bibinfo {author} {\bibfnamefont {B.}~\bibnamefont
  {Damski}},\ }\href
  {https://journals.aps.org/prl/abstract/10.1103/PhysRevLett.95.035701}
  {\bibfield  {journal} {\bibinfo  {journal} {Physical review letters}\
  }\textbf {\bibinfo {volume} {95}},\ \bibinfo {pages} {035701} (\bibinfo
  {year} {2005})}\BibitemShut {NoStop}%
\bibitem [{\citenamefont {Saito}\ \emph {et~al.}(2007)\citenamefont {Saito},
  \citenamefont {Kawaguchi},\ and\ \citenamefont {Ueda}}]{saito2007kibble}%
  \BibitemOpen
  \bibfield  {author} {\bibinfo {author} {\bibfnamefont {H.}~\bibnamefont
  {Saito}}, \bibinfo {author} {\bibfnamefont {Y.}~\bibnamefont {Kawaguchi}},\
  and\ \bibinfo {author} {\bibfnamefont {M.}~\bibnamefont {Ueda}},\ }\href
  {https://journals.aps.org/pra/abstract/10.1103/PhysRevA.76.043613} {\bibfield
   {journal} {\bibinfo  {journal} {Physical Review A}\ }\textbf {\bibinfo
  {volume} {76}},\ \bibinfo {pages} {043613} (\bibinfo {year}
  {2007})}\BibitemShut {NoStop}%
\bibitem [{\citenamefont {Nowak}\ and\ \citenamefont
  {Dziarmaga}(2021)}]{nowak2021quantum}%
  \BibitemOpen
  \bibfield  {author} {\bibinfo {author} {\bibfnamefont {R.~J.}\ \bibnamefont
  {Nowak}}\ and\ \bibinfo {author} {\bibfnamefont {J.}~\bibnamefont
  {Dziarmaga}},\ }\href
  {https://journals.aps.org/prb/abstract/10.1103/PhysRevB.104.075448}
  {\bibfield  {journal} {\bibinfo  {journal} {Physical Review B}\ }\textbf
  {\bibinfo {volume} {104}},\ \bibinfo {pages} {075448} (\bibinfo {year}
  {2021})}\BibitemShut {NoStop}%
\bibitem [{\citenamefont {Schmitt}\ \emph {et~al.}(2022)\citenamefont
  {Schmitt}, \citenamefont {Rams}, \citenamefont {Dziarmaga}, \citenamefont
  {Heyl},\ and\ \citenamefont {Zurek}}]{schmitt2022quantum}%
  \BibitemOpen
  \bibfield  {author} {\bibinfo {author} {\bibfnamefont {M.}~\bibnamefont
  {Schmitt}}, \bibinfo {author} {\bibfnamefont {M.~M.}\ \bibnamefont {Rams}},
  \bibinfo {author} {\bibfnamefont {J.}~\bibnamefont {Dziarmaga}}, \bibinfo
  {author} {\bibfnamefont {M.}~\bibnamefont {Heyl}},\ and\ \bibinfo {author}
  {\bibfnamefont {W.~H.}\ \bibnamefont {Zurek}},\ }\href
  {https://www.science.org/doi/full/10.1126/sciadv.abl6850} {\bibfield
  {journal} {\bibinfo  {journal} {Science Advances}\ }\textbf {\bibinfo
  {volume} {8}},\ \bibinfo {pages} {eabl6850} (\bibinfo {year}
  {2022})}\BibitemShut {NoStop}%
\bibitem [{\citenamefont {Naz{\'e}}\ \emph
  {et~al.}(2022{\natexlab{a}})\citenamefont {Naz{\'e}}, \citenamefont
  {Bonan{\c{c}}a},\ and\ \citenamefont {Deffner}}]{naze2022kibble}%
  \BibitemOpen
  \bibfield  {author} {\bibinfo {author} {\bibfnamefont {P.}~\bibnamefont
  {Naz{\'e}}}, \bibinfo {author} {\bibfnamefont {M.~V.~S.}\ \bibnamefont
  {Bonan{\c{c}}a}},\ and\ \bibinfo {author} {\bibfnamefont {S.}~\bibnamefont
  {Deffner}},\ }\href {https://www.mdpi.com/1099-4300/24/5/666} {\bibfield
  {journal} {\bibinfo  {journal} {Entropy}\ }\textbf {\bibinfo {volume} {24}},\
  \bibinfo {pages} {666} (\bibinfo {year} {2022}{\natexlab{a}})}\BibitemShut
  {NoStop}%
\bibitem [{\citenamefont {Kubo}\ \emph {et~al.}(2012)\citenamefont {Kubo},
  \citenamefont {Toda},\ and\ \citenamefont {Hashitsume}}]{kubo2012}%
  \BibitemOpen
  \bibfield  {author} {\bibinfo {author} {\bibfnamefont {R.}~\bibnamefont
  {Kubo}}, \bibinfo {author} {\bibfnamefont {M.}~\bibnamefont {Toda}},\ and\
  \bibinfo {author} {\bibfnamefont {N.}~\bibnamefont {Hashitsume}},\
  }\href@noop {} {\emph {\bibinfo {title} {Statistical physics II:
  nonequilibrium statistical mechanics}}},\ Vol.~\bibinfo {volume} {31}\
  (\bibinfo  {publisher} {Springer Science \& Business Media},\ \bibinfo {year}
  {2012})\BibitemShut {NoStop}%
\bibitem [{\citenamefont {Naz{\'e}}\ \emph
  {et~al.}(2022{\natexlab{b}})\citenamefont {Naz{\'e}}, \citenamefont
  {Deffner},\ and\ \citenamefont {Bonan{\c{c}}a}}]{naze2022optimal}%
  \BibitemOpen
  \bibfield  {author} {\bibinfo {author} {\bibfnamefont {P.}~\bibnamefont
  {Naz{\'e}}}, \bibinfo {author} {\bibfnamefont {S.}~\bibnamefont {Deffner}},\
  and\ \bibinfo {author} {\bibfnamefont {M.~V.~S.}\ \bibnamefont
  {Bonan{\c{c}}a}},\ }\href
  {https://iopscience.iop.org/article/10.1088/2399-6528/ac871d/meta} {\bibfield
   {journal} {\bibinfo  {journal} {Journal of Physics Communications}\ }\textbf
  {\bibinfo {volume} {6}},\ \bibinfo {pages} {083001} (\bibinfo {year}
  {2022}{\natexlab{b}})}\BibitemShut {NoStop}%
\bibitem [{\citenamefont {Deffner}(2017)}]{deffner2017}%
  \BibitemOpen
  \bibfield  {author} {\bibinfo {author} {\bibfnamefont {S.}~\bibnamefont
  {Deffner}},\ }\href {https://doi.org/10.1103/PhysRevE.96.052125} {\bibfield
  {journal} {\bibinfo  {journal} {Phys. Rev. E}\ }\textbf {\bibinfo {volume}
  {96}},\ \bibinfo {pages} {052125} (\bibinfo {year} {2017})}\BibitemShut
  {NoStop}%
\end{thebibliography}%
\bibliographystyle{apsrev4-2}

\end{document}